\documentclass[a4paper,11pt]{article}
\usepackage{pos}

\title{Hubble-induced phase transitions in the Standard Model and beyond}

\author*[a]{Javier Rubio}
\author[b]{Giorgio Laverda}

\affiliation[a]{Departamento de Física Teórica and Instituto de Física de Partículas y del Cosmos (IPARCOS-UCM), Universidad Complutense de Madrid, 28040 
Madrid, Spain} 

\affiliation[b]{Centro de Astrof\'{\i}sica e Gravita\c c\~ao  - CENTRA,
Departamento de F\'{\i}sica, Instituto Superior T\'ecnico - IST,
Universidade de Lisboa - UL,
Av. Rovisco Pais 1, 1049-001 Lisboa, Portugal} 

\emailAdd{javier.rubio@ucm.es}
\emailAdd{giorgio.laverda@tecnico.ulisboa.pt}
\abstract{We review the dynamics of spectator scalar fields non-minimally coupled to gravity in the post-inflationary Universe, with particular emphasis on scenarios where the end of inflation is followed by a period of kination. In this context, the evolution of the Ricci scalar can lead to the spontaneous breaking of discrete or continuous symmetries through tachyonic instabilities. These Hubble-induced phase transitions may cause the amplification of field fluctuations, the formation of transient topological defects, and an efficient energy transfer into relativistic degrees of freedom. We analyze this general mechanism and its cosmological consequences, including (re)heating and gravitational wave production. As a concrete realization, we discuss the postinflationary evolution of the Standard Model Higgs, examining the interplay between curvature effects, vacuum stability, and non-perturbative dynamics. This framework provides a minimal and predictive connection between high-energy physics and the early Universe, with potential observational signatures.}
\FullConference{
}

\begin{document}
\maketitle
\section{Introduction}

Early-Universe phase transitions can be driven not only by high temperatures but also by the background expansion rate. In particular, a Hubble-induced phase transition (HIPT) refers to a spontaneous symmetry breaking triggered by the time-dependence of the Hubble parameter in the aftermath of inflation \cite{Laverda:2023uqv, Laverda:2024qjt, Bettoni:2019dcw, Bettoni:2021zhq, Bettoni:2018utf, Bettoni:2018pbl, Mantziris:2024uzz, Kierkla:2023uzo}. In such scenarios, a spectator scalar field non-minimally coupled to gravity acquires a time-varying effective squared mass proportional to the Ricci scalar. As the Universe evolves from inflation to the post-inflationary epoch, this effective mass can transiently turn negative, inducing a tachyonic instability that drives the spectator field towards a new vacuum expectation value far away from the original symmetric state. 

\begin{figure}
    \centering
    \includegraphics[width=0.65\textwidth]{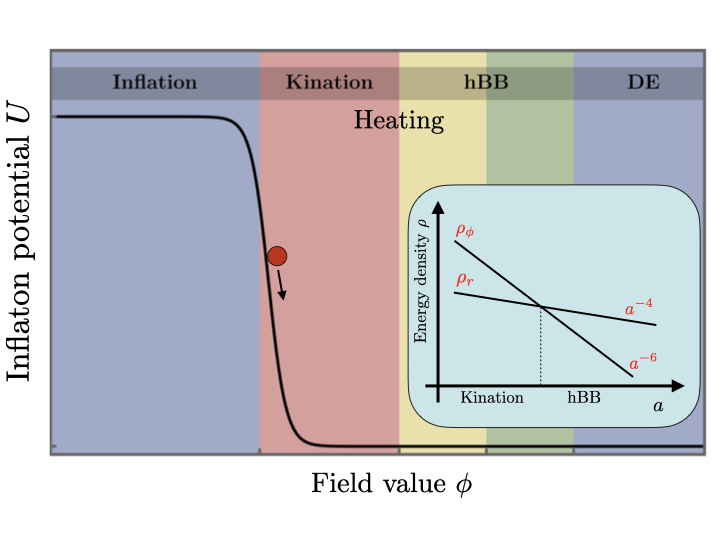}
    \caption{Illustration of the inflaton dynamics in quintessential inflation scenarios. The potential $U(\phi)$ features a steep decline at the end of inflation, followed by a kinetic-dominated phase (kination), and eventually transitions into radiation domination (hBB), matter domination, and late-time dark energy domination (DE). The inset highlights the energy densities of the inflaton ($\rho_\phi$) and radiation ($\rho_r$) as functions of the scale factor $a$, showcasing the delayed onset of radiation domination due to the faster dilution of kinetic energy ($\propto a^{-6}$) as compared to radiation ($\propto a^{-4}$).}
    \label{fig:inflaton_potential_evolution}
\end{figure}

HIPT mechanisms are especially compelling in models where inflation is not followed by a radiation-dominated phase, but by a period of kination — an epoch where the energy density of the Universe is dominated by the kinetic energy of the inflaton. Such a stiff phase is naturally realized in quintessential inflation models \cite{Bettoni:2021zhq} and variable gravity scenarios \cite{Wetterich:1987fm,Wetterich:1994bg,Rubio:2017gty}, cf.~Fig.~\ref{fig:inflaton_potential_evolution}. In this context, conventional (re)heating mechanisms \cite{Barman:2025lvk} relying on the complete decay of the inflaton condensate are typically suppressed, as this field must remain weakly coupled to avoid disrupting its late-time dynamics. Consequently, alternative routes to repopulating the Universe with radiation must be considered. In this regard, non-minimally coupled spectator fields provide a minimal and predictive framework, allowing not only for a symmetry breaking triggered by curvature effects, but also  for a sufficient energy transfer into relativistic degrees of freedom, thereby (re)heating the Universe in a purely gravitational fashion \cite{Bettoni:2021zhq,Laverda:2023uqv}.

Beyond their role in (re)heating, HIPTs offer a remarkably rich phenomenology. The tachyonic instability that follows the sign flip in the effective mass leads to a rapid amplification of field fluctuations, giving rise to inhomogeneous field configurations characterized by the emergence of transient topological defects — most notably, domain walls or cosmic strings, depending on the symmetry of the field \cite{Bettoni:2018pbl,Bettoni:2019dcw}. These defects are not permanent: they form and decay within a short period, avoiding the standard problems associated with long-lived relics. However, while they exist, they can significantly influence the dynamics of the early Universe. The oscillations and interactions of these defects, as well as the subsequent non-linear evolution of the field, inject anisotropic stresses into the spacetime, sourcing a stochastic background of gravitational waves and offering a unique window into physics at energy scales far beyond the reach of terrestrial experiments  \cite{Bettoni:2024ixe}.

A particularly well-motivated application of this mechanism arises when the spectator field is identified with the Standard Model Higgs \cite{Laverda:2024qjt,Laverda:2025pmg}. The Higgs field, due to its scalar nature and its unavoidable presence in the early Universe, is naturally sensitive to curvature effects. Moreover, the Higgs potential is known to exhibit metastability at high energies: depending on the top quark mass and other parameters, the electroweak vacuum may be separated from a deeper true vacuum by a potential barrier. During inflation, a non-minimal coupling to gravity can stabilize the Higgs and protect against vacuum decay. However, the same coupling can become dangerous in the post-inflationary phase, where the Ricci scalar becomes negative. This raises the possibility that the Higgs could undergo a curvature-induced transition toward the instability region, threatening electroweak vacuum stability. This interplay imposes tight constraints on the inflationary scale and the size of the Higgs–curvature coupling: (re)heating must occur efficiently enough to trigger radiation domination, but not so violently as to push the Higgs beyond the potential barrier \cite{Laverda:2024qjt,Laverda:2025pmg}

This article is structured as follows. In Section \ref{sec:scenario}, we introduce a simplified toy model featuring a non-minimally coupled scalar field and analyze its dynamics in the post-inflationary Universe. Section \ref{sec:PT} and \ref{sec;NL} focus respectively on the linear and non-linear regimes of the phase transition, including defect formation and inhomogeneity growth. In Section \ref{sec:hBB}, we examine how these effects drive the onset of radiation domination, quantifying the efficiency of (re)heating. Section \ref{sec:grav} discusses the associated gravitational wave production and its observational signatures. Section \ref{sec:Higgs} turns to the specific case of the Standard Model Higgs, outlining the constraints from vacuum stability and presenting the range of parameters compatible with successful Higgs reheating. We conclude in Section \ref{sec:conc} with a summary of our findings and a discussion of possible extensions and observational prospects.

\section{The simpler scenario}\label{sec:scenario}

Consider a non-minimally coupled and \textit{energetically subdominant spectator} field $\chi$ with a simple $Z_2$ symmetry $\chi \to -\chi$ and Lagrangian density,\footnote{If the broken symmetry were continuous rather than $Z_2$, analogous defects like cosmic strings or textures could form depending on the vacuum manifold \cite{Bettoni:2018pbl}.  Here, we mainly consider the $Z_2$ case for concreteness, but the general phenomena would be similar for other symmetries.}
\begin{equation}\label{eq:lagchi} S_\chi=\int d^4 x \sqrt{-g} \left[-\frac{1}{2}g^{\mu\nu}\partial_\mu\chi \partial_\mu \chi-\frac{1}{2}\xi R \chi^2-\frac14 \lambda \chi^4 \right]\,, 
\end{equation}
where we have deliberately excluded a potential mass term, under the assumption that it will have a negligible influence on the following analysis  \cite{Bettoni:2018utf,Bettoni:2018pbl,Bettoni:2019dcw}. In this expression, $R$ denotes the Ricci scalar associated to the metric $g_{\mu\nu}$, and $\xi$ and $\lambda$ are dimensionless constants, both taken to be strictly positive in what follows, unless otherwise stated. The dynamics of the $\chi$ field is characterized by a generalized Klein–Gordon equation 
\begin{equation}\label{eq:eq_chicov}
 \Box\, \chi -\xi R\chi-\lambda \chi^3=0\,,
\end{equation}
and a stress-energy tensor 
\begin{equation}
T^{(\chi)}_{\mu\nu}  = 
  D_{\mu} \chi  D_{\nu} \chi - g_{\mu\nu} \left[
   \frac{1}{2}D^\rho \chi D_\rho\chi    
  + \frac{\lambda}{4} \, \chi^4 \right]  +\xi \Big[ G_{\mu\nu} 
  + \left(g_{\mu\nu} \Box -D_{\mu} \, D_{\nu}\right)
 \Big] \chi^2,    
  \label{eq:emt}        
\end{equation}
with $D_\mu$ indicating the covariant derivative, $G_{\mu\nu}$ the Einstein tensor, and $\Box = D^\lambda D_\lambda$ the d'Alembertian operator. Note that $T^{(\chi)}_{\mu\nu}u^\mu u^\nu$ is not necessarily positive definite for every timelike vector $u^\mu$, allowing for a potential violation of the weak energy condition within this particular sector \cite{Ford:1987de,Ford:2000xg,Bekenstein:1975ww,Flanagan:1996gw}.

For a spatially flat Friedmann–Lemaître–Robertson–Walker (FLRW) Universe with metric $g_{\mu\nu} = \text{diag}(-1, a^2(t)\delta_{ij})$, the Ricci scalar in Eq.~\eqref{eq:eq_chicov} takes the form $R=3(1-3w) H^2$, with $H = \dot{a}/a$ the Hubble parameter, the overdots indicating time derivatives, and $w$ the background equation of state, implicitly identified with that of the dominant inflaton component $\phi$ in the following developments, $w\simeq w_\phi$. During inflation ($w \approx -1$) the Ricci scalar $R = 12H^2$ is positive and large, inducing a sizable and positive effective mass-squared for the spectator field, $M^2_{\rm eff} \sim \xi R$, that stabilizes it at the origin. An important consequence of this stabilization is the absence of isocurvature perturbations \cite{Bettoni:2018utf,Laverda:2024qjt, Herranen:2014cua, Herranen:2015ima, Kohri:2016wof}.~\footnote{Isocurvature modes arise when a light field has quantum fluctuations during inflation that are independent of the curvature (density) perturbations. Such isocurvature signals are tightly constrained by cosmic microwave background observations, so it is crucial that they remain suppressed.} Once inflation ends and the Universe enters a kinetic-dominated epoch ($w \approx 1$), the Ricci scalar becomes negative ($R<0$), effectively making the effective mass term tachyonic (${M^2_{\rm eff}}<0$). This Hubble-induced contribution triggers the spontaneous symmetry breaking of the field’s $Z_2$ symmetry and drives $\chi$ away from the false vacuum at $\chi=0$ toward new minima at large field values. In essence, the rapid drop and sign flip of $R$ acts like a quench, turning what was a symmetric potential into a Mexican-hat potential with emerging degenerate minima. While one might naively picture the field as evolving homogeneously to one of the new minima \cite{Opferkuch:2019zbd,Figueroa:2016dsc,Fairbairn:2018bsw}, in reality the process is highly inhomogeneous \cite{Bettoni:2019dcw}. The tachyonic instability amplifies exponentially long-wavelength fluctuations of $\chi$ that had been held in check so far by the Hubble friction. In fact, before $\chi$ can even settle into the minima, different regions of the Universe begin to fall into different sides of the potential. Thus, instead of a uniform condensate, a complex field configuration develops, characterized by large oscillating domains of $\chi$, as discussed in detail in the next section. 

\section{Phase transition dynamics and defect formation}\label{sec:PT}

Right after the $Z_2$ symmetry is broken by Hubble-induced effects, the dynamics described by the action \eqref{eq:eq_chicov} effectively reduces to that of a free spectator scalar field whose mass varies with time. This simplification renders the system analytically tractable, allowing us to characterize the typical shape and scales of the domains at the time of formation, where non-linearities can no longer be ignored. Modelling the transition from inflation to kination as $a= a_{\rm kin} \left[1+3 H_{\rm kin} \left(t-t_{\rm kin}\right)\right]^{1/3}$ and $H=H_{\rm kin}(1+3H_{\rm kin}(t-t_{\rm kin}))^{-1}$, with $a_{\rm kin}$ and $H_{\rm kin}$ the values of the scale factor and the Hubble rate at the transition time $t_{\rm kin}$, and introducing rescaled variables $Y\equiv a {\chi}/(a_{\rm kin}\chi_*)$, $y \equiv \, a_{\rm kin} \chi_* \,\vec x$ and $z \equiv \, a_{\rm kin}\chi_* \, \tau$, with $\chi_*\equiv \sqrt{6\xi} H_{\rm kin}$,
the Fourier-transformed equation of motion \eqref{eq:lagchi} in this regime becomes that of a free harmonic oscillator \cite{Bettoni:2019dcw}
\begin{equation}\label{EOModes}
Y''_{\vec\kappa}(z)+ \omega_{\kappa}^2(z) Y_{\vec\kappa}(z)=0\,,
\end{equation}
with time-dependent frequency $\omega_{\kappa}^2(z)\equiv \kappa^2-M(z)^2$. The effective mass
\begin{equation} 
M^2(z)\equiv (1-6\xi)(\mathcal{H}^2+\mathcal{H}')=(4\nu^2-1)\mathcal{H}^2\,,
\end{equation}
in this expression depends the comoving Hubble parameter $\mathcal{H}(z)\equiv a'(z)/a(z)=1/{(2(z+\nu))}$ and a rescaled non-minimal coupling parameter  $\nu = \sqrt{3\xi/2}$. For $\nu>1
/2$, Eq.~\eqref{EOModes} displays a tachyonic instability during the kinetic-dominated regime, which amplifies not only the zero mode of the spectator field but also a range of subhorizon modes whose comoving momenta lie in the interval $\mathcal{H}(z) \lesssim \kappa \lesssim (4\nu^2-1)^{1/2}\mathcal{H}(z) $. All statistical information of the quantum field in this Gaussian setup is contained in the two-point correlators $\langle v^I_{\vec\kappa}(z), v^J_{\vec \kappa'}(z') \rangle = \Sigma^{IJ}_\kappa(z,z')\delta^3({\vec \kappa}+{\vec \kappa'})$ at equal or different times,
with  $v_{\vec \kappa} \equiv \left(\Pi_{\vec \kappa}(z),Y_{\vec\kappa}(z)\right)^T$, $Y_{\vec\kappa}(z)$ and $\Pi_{\vec \kappa}(z)$ canonical position and momentum operators fulfilling the standard equal-time commutation relation $\left[Y_{\vec\kappa}(z),\Pi_{\vec \kappa'}(z)\right]=i \delta^3(\vec \kappa + \vec \kappa')$ and the expectation value taken over the quantum state at the initial time. 
\begin{figure}
    \centering
    \includegraphics[scale=0.58]{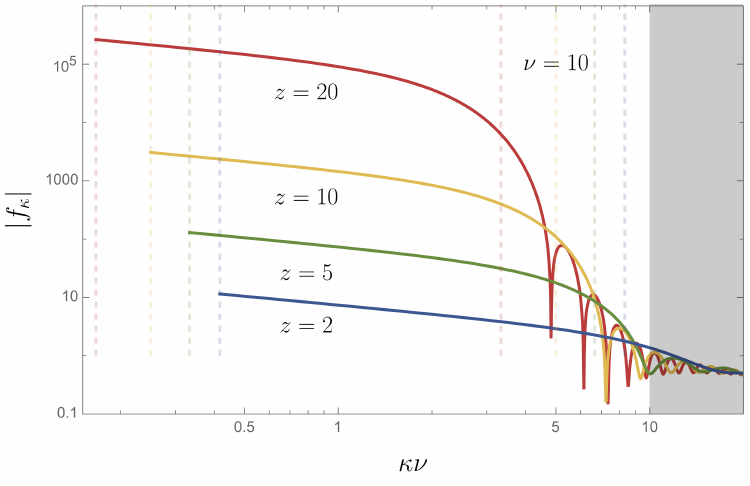}
    \includegraphics[scale=0.58]{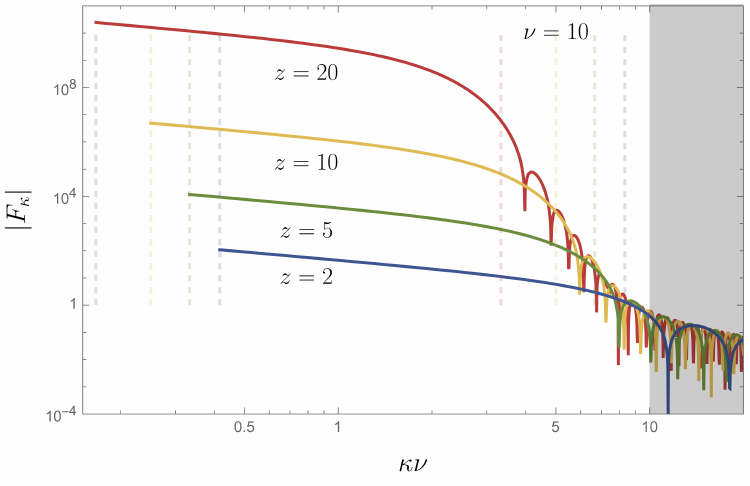}
    \caption{Amplitude evolution of the mode functions $\vert f_\kappa\vert$ (left) and $\vert F_\kappa\vert$ (right) at several conformal times $z$. The shaded region marks the range of momenta outside the initial amplification band, defined by $\kappa > \kappa(z=0)$. Dashed vertical lines indicate the boundaries of the time-dependent amplification band  for each corresponding time. Taken from \cite{Bettoni:2021qfs}.}
    \label{fig:fkFk}
\end{figure} 
Solving the evolution of this system in the Heisenberg picture, 
\begin{equation}\label{eq:Heise} 
\frac{d}{dz} \begin{pmatrix} \Pi_{\vec \kappa}(z) \vspace{2mm} \cr Y_{\vec\kappa}(z) \end{pmatrix}= \begin{pmatrix} 0 & -\omega^2_{\kappa}(z)\vspace{2mm}\cr 1 &0 \end{pmatrix} \begin{pmatrix}\Pi_{\vec\kappa}(z) \vspace{2mm} \cr Y_{\vec \kappa}(z) \end{pmatrix} \,,
\end{equation}
we obtain an equal-time covariance matrix 
\begin{equation}\label{Sigmaz} \Sigma_{\kappa}(z,z)= \begin{pmatrix} |g_{\kappa}(z)|^2 & F_\kappa(z) - \frac{i}{2}\vspace{3mm}\cr F_\kappa(z) + \frac{i}{2} & |f_k(z)|^2 \end{pmatrix}\,, 
\end{equation}
with $f_\kappa(z)$ satisfying the differential equation $f_\kappa '' + \omega^2_\kappa(z)f_\kappa = 0$, with initial data $f_\kappa(z_r)$ and $f'_\kappa(z_r)$, $g_\kappa(z) \equiv i f'_\kappa(z)$, and $F_\kappa(z)\equiv \frac12\,\langle \Pi_{\vec \kappa}(z)\, Y_{\vec \kappa}^\dagger(z) + Y_{\vec \kappa}(z) \Pi_{\vec \kappa}^\dagger(z)\rangle = {\rm Im}\, (f_\kappa^* g_\kappa)$ 
a phase-space indicator directly tied to the uncertainty relation
$
\Delta Y_\kappa^2\,\Delta \Pi_\kappa^2 = \vert F_\kappa(z)\vert^2 + 1/4 \geq 1/4 \vert \langle [Y_\kappa(z),\, \Pi^\dagger_\kappa(z)]\rangle\vert^2\,. 
$
When $|F_\kappa(z)| \gg 1$, the anticommutator greatly exceeds the canonical commutator, $\langle{Y_\kappa(z),\ \Pi_\kappa^\dagger(z)}\rangle \gg \langle|[ Y_\kappa(z), \Pi_\kappa^\dagger(z)]|\rangle = \hbar$, indicating that the system can be faithfully described by a classical stochastic field. As shown in Fig.~\ref{fig:fkFk}, this behaviour emerges shortly after the onset of the HIPT, thereby justifying the use of classical lattice simulations to study the system's subsequent evolution \cite{Bettoni:2019dcw}.

\section{Non-linear regime} \label{sec;NL}

\begin{figure}
    \centering
    \includegraphics[width=0.33\textwidth]{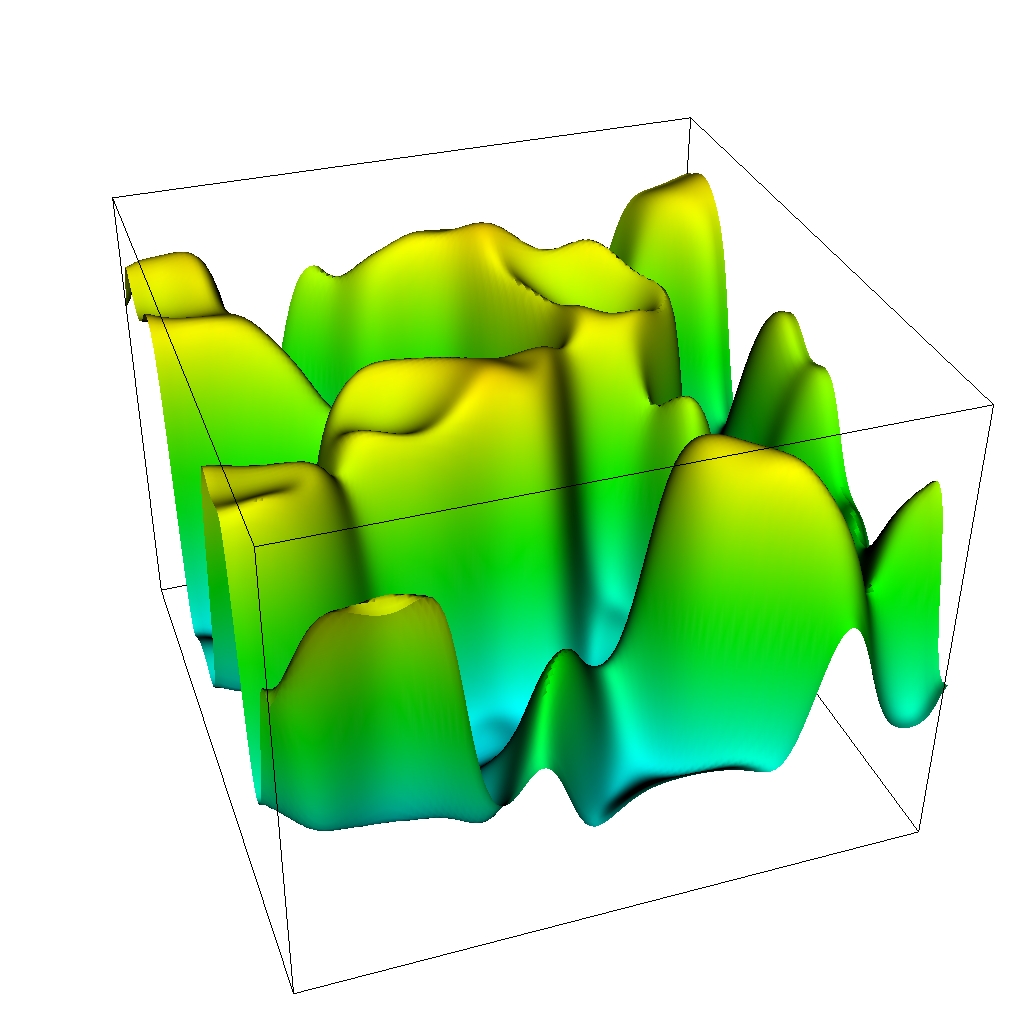}
    \includegraphics[width=0.293\textwidth]{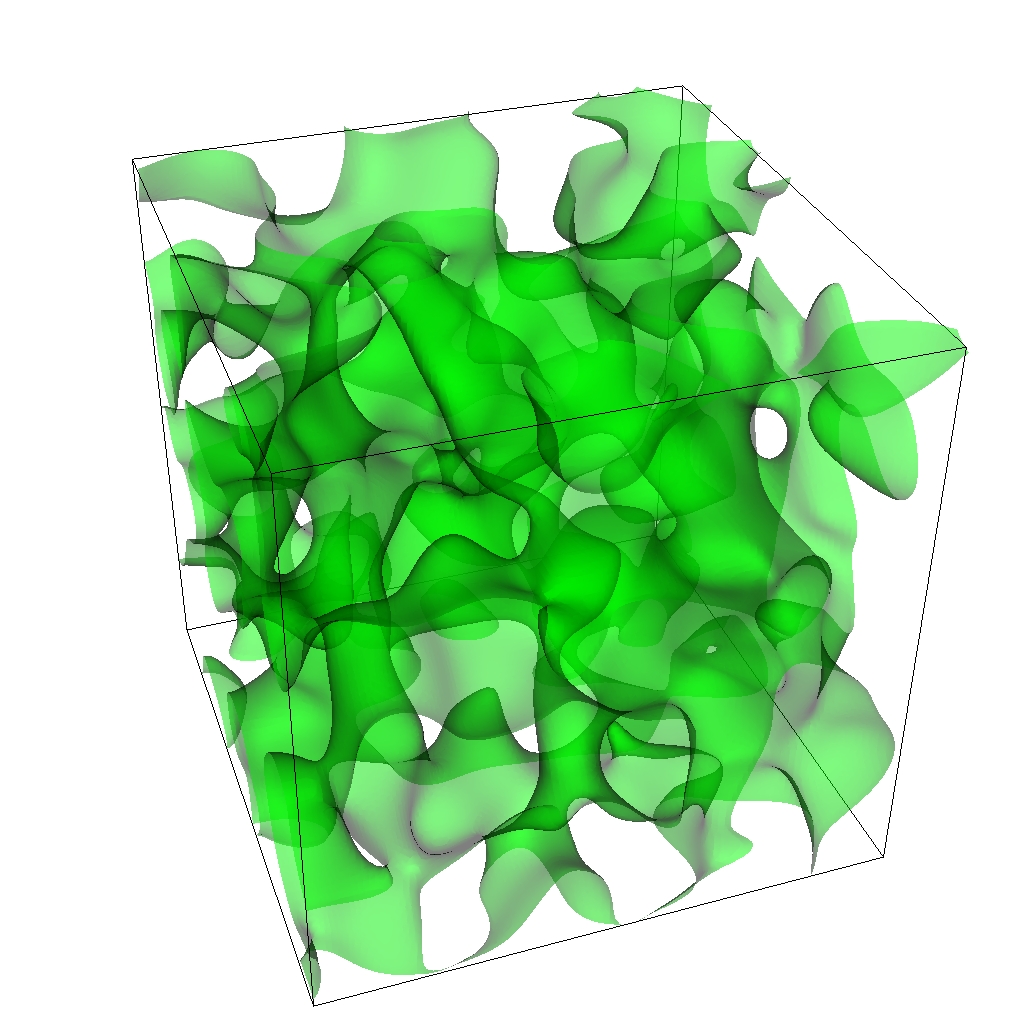}
    \includegraphics[width=0.363\textwidth]{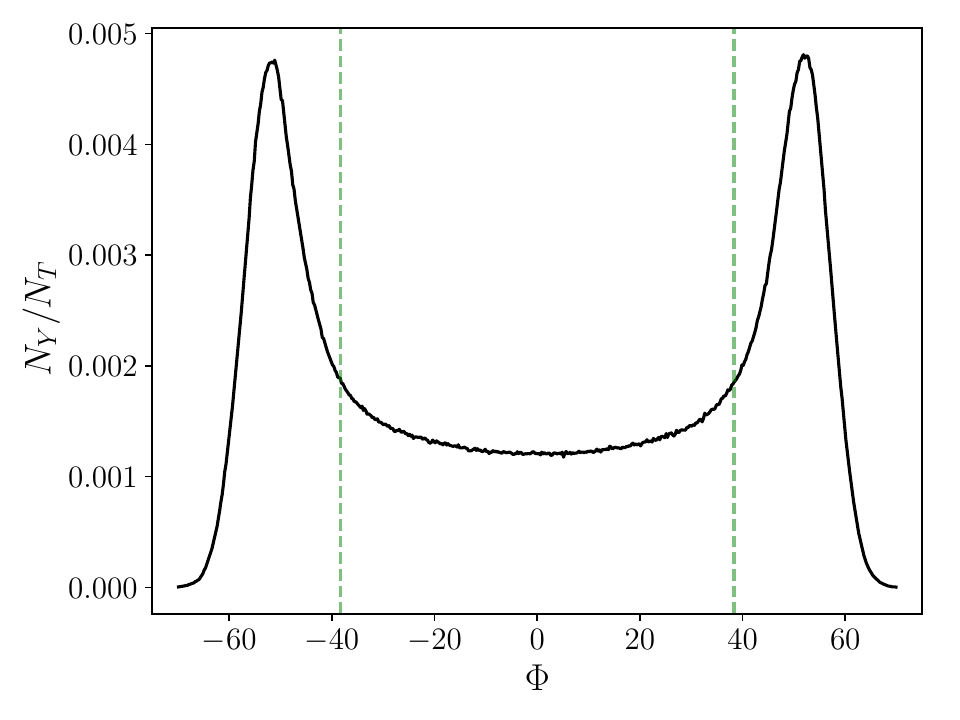}
    \caption{Formation of Hubble-induced domain walls illustrated from three complementary viewpoints for the benchmark point $\nu=10$, $\lambda=10^{-4}$. Left: 2D slice showing the field profile across a section of the simulation volume. Center: 3D visualization of domain wall structures identified by zero-field crossings. Right: Field probability distribution highlighting the emergence of a bimodal structure corresponding to the degenerate minima of the effective potential. All snapshots correspond to time $z=5$. Adapted from \cite{Bettoni:2021zhq}.}
    \label{fig:domainwall_formation}
\end{figure}

Lattice simulations of HIPTs vividly confirm the previous analytical picture, showing that very soon after the transition, the field configuration consists of large coherent domains, divided by a network of domain walls where $\chi \approx 0$ \cite{Bettoni:2021zhq,Laverda:2023uqv}. The formation of Hubble-induced domain walls is illustrated in Fig.~\ref{fig:domainwall_formation}, which captures three complementary results from lattice simulations at a representative time shortly after the transition ($z=5$). The left panel shows a 2D slice of the simulation volume, clearly revealing extended domains where the field $\chi$ has settled near one of the two minima, separated by sharp interfaces where $\chi \approx 0$. These interfaces correspond to domain walls. The center panel provides a 3D visualization of the same structures, highlighting the intricate and interconnected nature of the wall network formed via zero-field crossings. Finally, the right panel presents the field probability distribution, showing the emergence of a bimodal structure that reflects the spontaneous symmetry breaking and the formation of degenerate minima. Interestingly, as the field oscillates and evolves, these domain walls do not remain static, but rather appear and disappear periodically. In other words, the oscillations of $\chi$ can temporarily restore $\chi\approx0$ in specific regions, effectively “melting” the walls, which then reform slightly shifted after each oscillation. The network is therefore dynamical – domain walls pulsate and move as the field oscillations continue.  The presence of these defects is one of the hallmark features of the HIPT, distinguishing it from a purely homogeneous symmetry-breaking scenario.

Although pulsating domain walls form, they are not permanent in this framework. As the Hubble rate $H(t)$ drops further, the curvature term $\xi R\chi^2$ becomes smaller in magnitude and the Hubble-induced tachyonic mass term eventually becomes negligible. In other words, the effective symmetry breaking is a transient phenomenon: after some time, the two minima begin to approach each other and the vacuum migrates back toward $\chi=0$ as the true ground state. When this happens, the domain walls separating the false vacua cannot persist – they annihilate each other as the field in all regions relaxes towards a common vacuum state \cite{Bettoni:2021zhq,Laverda:2023uqv}. In fact, the HIPT provides a natural solution to the usual domain wall problem: the walls are created, but they are “not forever” – they are eliminated by the time radiation domination is established \cite{Bettoni:2019dcw}. The outcome is a Universe free of domain walls, but not before these defects play a significant role in the intermediate dynamics, as we will see in what follows.  

\section{Ricci reheating and the onset of radiation domination}\label{sec:hBB}

\begin{figure}[t]
    \centering
    \includegraphics[width=0.7\textwidth]{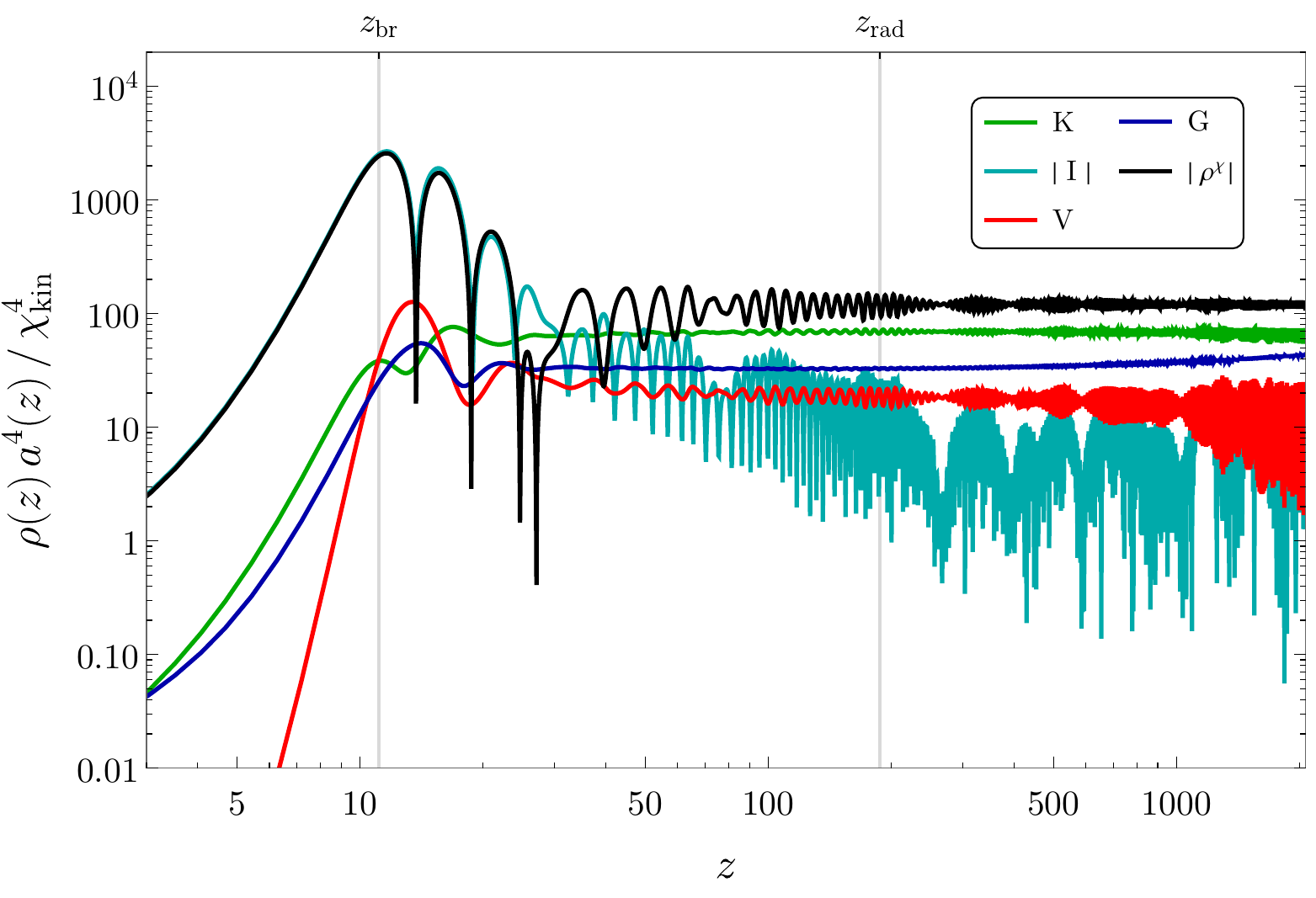}
    \caption{
Time evolution of the spatially averaged energy components—kinetic ($K$), gradient ($G$), potential ($V$), and interaction ($I$)—contributing to the total scalar field energy density $\rho^{\chi}$, for a representative case with $\nu = 10$ and $\lambda = 10^{-4}$. To emphasize the approach toward a radiation-like equation of state, each component is rescaled by a factor of $a^4$. The absolute values of the interaction term and total energy are shown, as these quantities can become temporarily negative during the non-linear evolution. The backreaction onset ($z_{\rm br}$) and the virialization point ($z_{\rm rad}$) for this configuration are marked along the top axis. Taken from \cite{Bettoni:2021zhq}. }
    \label{fig:single_scalar_energies}
\end{figure}
 \begin{figure}[t]
        \centering
        \includegraphics[width=0.75\textwidth]{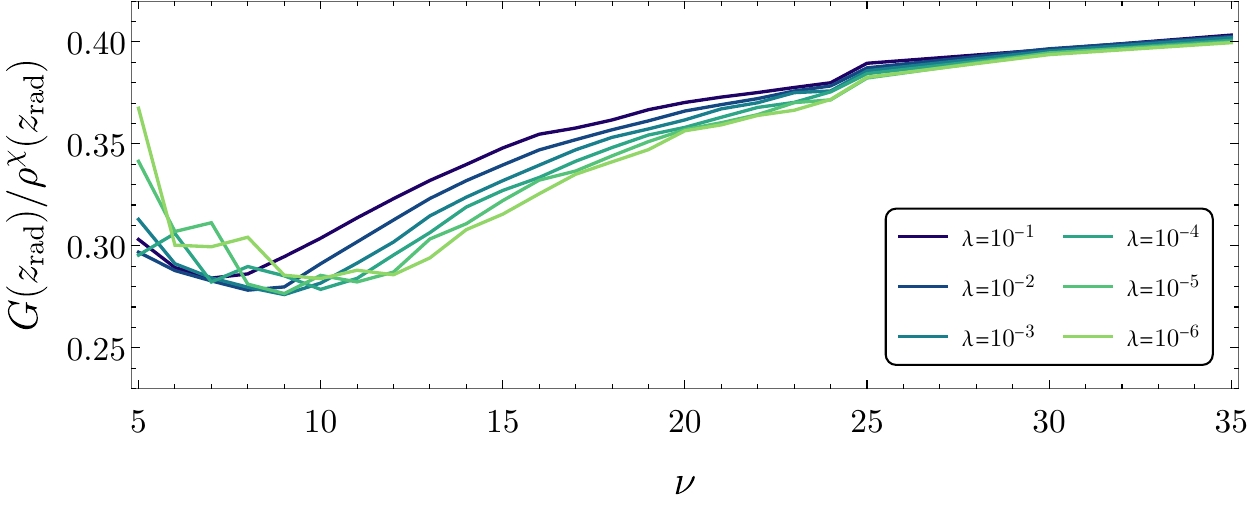}
        \caption{Fraction of the total spectator field energy density carried by spatial gradients at the onset of radiation domination ($z_{\rm rad}$), shown as a function of the parameters $\nu$ and $\lambda$. The emergence of a radiation-like equation of state hinges on the significant contribution of gradient energy, underscoring the non-negligible role of inhomogeneities in the post-instability dynamics. Taken from \cite{Laverda:2023uqv}. }
    \label{fig:grad_kin_ratio}
\end{figure} 
 \begin{figure}[t]
        \centering
        \includegraphics[width=0.8\textwidth]{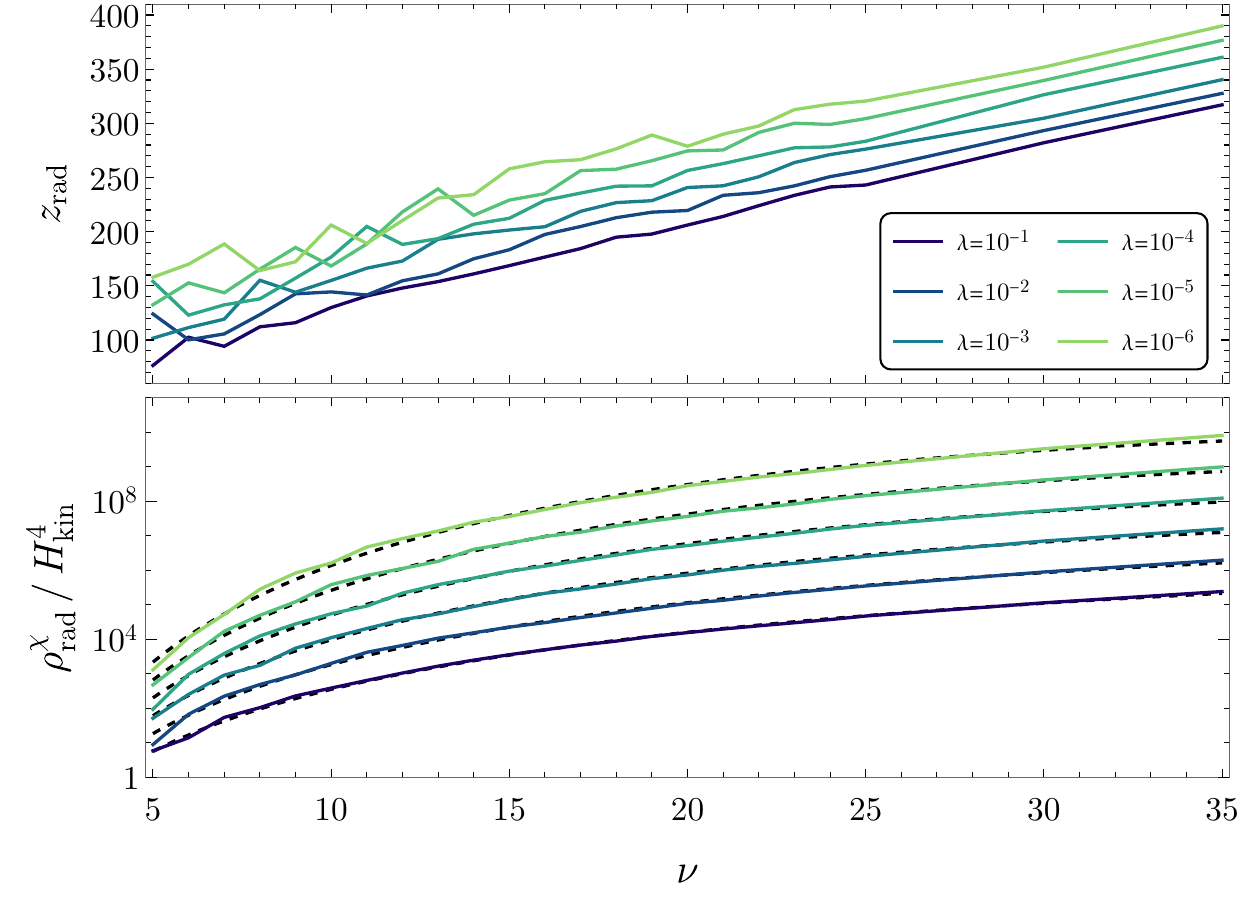}
        \caption{Dependence of key post-instability observables on the model parameters $\nu$ and $\lambda$. The upper panel shows how the virialization timescale $z_{\text{rad}}$ varies across parameter space, while the lower panel displays the corresponding total energy density $\rho^{\chi}_{\text{rad}}$ once virialization is achieved. Colored solid lines represent the results extracted from a full ensemble of lattice simulations. In the lower plot, black dashed lines depict the analytic fit given by Eq.~\eqref{eq:fit_rho_rad}, illustrating its agreement with the numerical data. Taken from \cite{Laverda:2023uqv}}
    \label{fig:radiation_total}
\end{figure}
\begin{figure}
    \centering
    \includegraphics[scale=0.55]{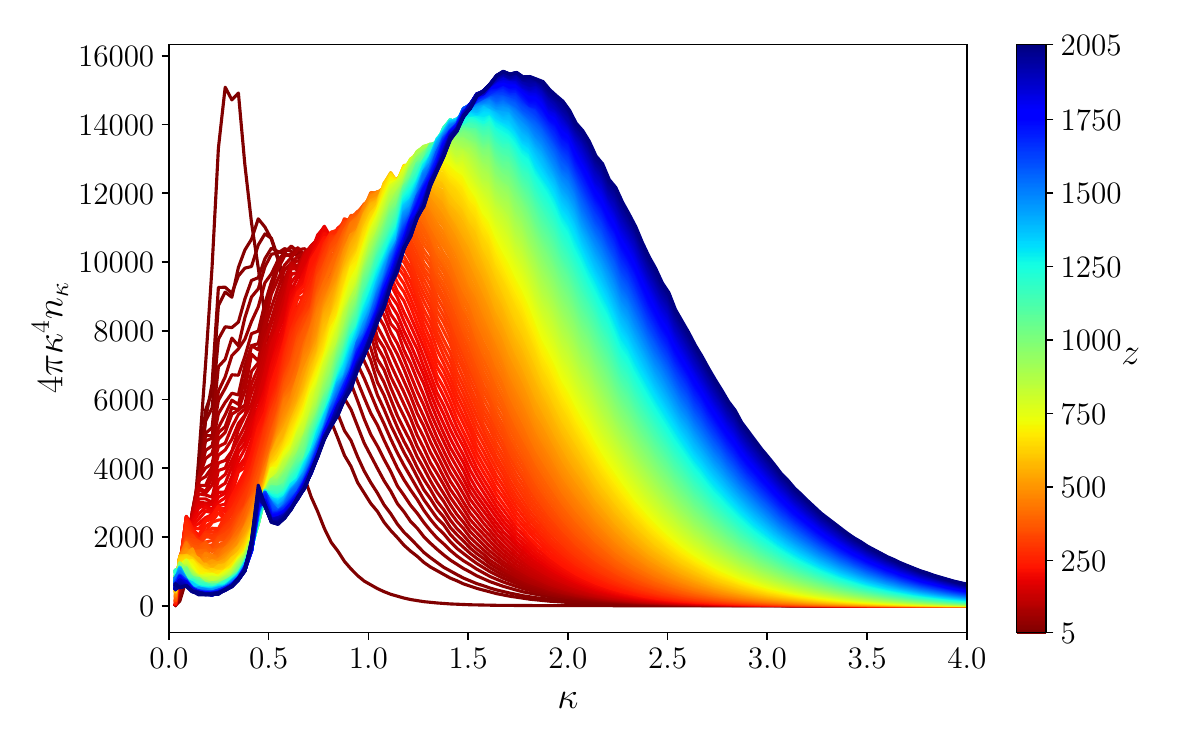}
    \caption{Momentum-space distribution of excitation modes for the benchmark configuration $\nu = 10$, $\lambda = 10^{-4}$. The plot displays the evolution of occupation numbers, highlighting how energy populates various Fourier modes over time. Distinct colors represent successive stages of the field's evolution, capturing the build-up and redistribution of power across momentum scales. Taken from \cite{Bettoni:2021zhq}}
    \label{fig:occnum}
\end{figure}

An immediate consequence of the Hubble-induced transition is an explosive production of field excitations – effectively, particle production out of the vacuum energy of $\chi$. The large-amplitude oscillations of the field, together with the inhomogeneous patches and domains, lead to a significant transfer of energy from the homogeneous mode into gradient energy and ultimately into relativistic particles \cite{Bettoni:2021zhq,Laverda:2023uqv,Figueroa:2024asq}. To understand how the Universe transitions toward a radiation-like phase, it is instructive to examine the time evolution of the different components of the field’s energy density. Fig.~\ref{fig:single_scalar_energies} displays the kinetic, gradient, potential, and interaction energies (all rescaled by $a^4$ to highlight the evolution of the equation of state) for a benchmark case. One can see that following the symmetry breaking, energy is rapidly redistributed from the homogeneous mode into gradients and interactions, signalling the fragmentation of the field. Notably, the gradient energy fraction grows until it roughly equals the kinetic component. The crucial result is that the equation of state of the $\chi$ field, $w_\chi = p_\chi/\rho_\chi$, evolves toward $1/3$, the equation of state of radiation. This end-state is largely independent of the detailed shape of the potential or the self-coupling strength, i.e. for all practical purposes, the system forgets the details of the symmetry breaking potential and approaches a universal radiation-like behaviour \cite{Bettoni:2021zhq}. This picture is further supported by Fig.~\ref{fig:grad_kin_ratio}, which quantifies the relative importance of gradient energy across parameter space at the moment radiation domination begins \cite{Laverda:2023uqv}. The results confirm that spatial gradients contribute a substantial fraction of the total energy, especially in regions of parameter space where the self-coupling $\lambda$ is small. This underscores the crucial role of inhomogeneities in the (re)heating dynamics: it is not just the coherent oscillations of $\chi$, but the development of structure and gradients that drives the transition to $w_\chi \approx 1/3$. Note in particular that without significant gradient contributions, the system would fail to virialize and behave as radiation for a generic potential $\chi^n$ \cite{Podolsky:2005bw,Lozanov:2016hid,Bettoni:2021zhq}. 

The time at which the system virializes and radiation domination begins is denoted by $z_{\rm rad}$. Based on a wide scan of lattice simulations (cf.~Fig.~\ref{fig:radiation_total}), the virialization time across parameter space is well described by the fitting formula \cite{Laverda:2023uqv} 
\begin{equation}\label{zradfit}
    z_{\text{rad}}(\lambda, \nu) = \gamma_1 + \gamma_2 \, \nu \,, 
\end{equation}
where
\begin{equation}
    \gamma_1(\lambda) = 33.63 + 15.02 \, n - 0.22 \, n^2 \,,\hspace{5mm}
    \gamma_2(\lambda) = 7.91 - 0.01 \, n + 0.02 \, n^2 \,,  \hspace{5mm} n=-\log (\lambda)\,.
\end{equation}
Similarly, the total energy density of the spectator field at virialization is well fitted by \cite{Laverda:2023uqv}
\begin{equation}
    \rho^{\chi}_{\text{rad}}(\lambda, \nu) = 16 H^4_{\rm kin} \, \exp\left({\delta_1 + \delta_2 \, \nu} +{\delta_3}\ln \nu\right) \,, 
\label{eq:fit_rho_rad}
\end{equation}  
with coefficients
\begin{equation}
    \delta_1(\lambda) = -11.10 - 0.06 \, n  \,,\hspace{5mm}
    \delta_2(\lambda) = -0.04 - 0.03 \, n \,, \hspace{5mm}
    \delta_3(\lambda) = 5.62 + 0.87 \, n \,. \label{coeff_rho_rad}
\end{equation}
From this, one defines the (re)heating efficiency $\Theta_{\rm ht}$ \cite{Rubio:2017gty} as the ratio of energy densities of the spectator and inflaton fields at the time of radiation domination, namely \cite{Laverda:2023uqv}
\begin{equation} \label{eq:fittingTheta}
    \Theta_{\text{ht}}(\lambda, \nu) \equiv \frac{\rho^{\chi}(a_{\rm rad})}{\rho^{\phi}(a_{\rm rad})} = \frac{16}{3}\left(\frac{H_{\rm kin}}{M_{P}}\right)^2  \left(1+\frac{\gamma_1 + \gamma_2 \, \nu}{\nu}\right)^3 \exp\left({\delta_1 + \delta_2 \, \nu} +\delta_3 \ln \nu \right)  \,,
\end{equation}
with $\rho^{\phi}(a)=3 M^2_{P}H^2_{\rm kin}(a/a_{\rm kin})^{-6}$. Crucially, these parametric expressions allow one to bypass expensive simulations for each new parameter choice, offering predictive control over the (re)heating dynamics across a wide range of $\lambda$ and $\xi$ \cite{Laverda:2023uqv}.

Meanwhile, the original inflaton field is still in a kination regime with equation of state $w \approx 1$. A crucial point is that a stiff fluid ($w=1$) redshifts away faster than radiation: $\rho_{\rm stiff}\propto a^{-6}$ whereas $\rho_{\rm rad}\propto a^{-4}$. Therefore, any small amount of radiation present will eventually overtake the stiff-component as the dominant energy. In our case, the Ricci-driven particle production by $\chi$ provides exactly such a radiation component. Even if $\chi$ starts subdominant compared to the inflaton’s kinetic energy, the combination of its internal dynamics and cosmological redshift ensures that radiation takes over after some time.  In short, the Universe naturally makes a transition to radiation domination: the HIPT (re)heats the Universe by converting vacuum and kinetic energy into relativistic particles. One subtlety is that, as illustrated in Fig.~\ref{fig:occnum},  the produced particle spectrum is initially far from equilibrium and therefore non-thermal \cite{Bettoni:2021zhq}.  This means concepts like a well-defined temperature immediately after the transition are tricky. Thermalization to a true radiation bath may require further scattering processes after $w=1/3$ is reached. Nevertheless, the equation of state and energy density behaviour are as if a radiation fluid were present, which is what matters for the expansion history.

Assuming thermalization by that time, the (re)heating efficiency \eqref{eq:fittingTheta} can be translated into an instantaneous temperature at the onset of radiation domination using the standard relation \cite{Laverda:2023uqv}
\begin{equation}
T_{\rm ht} =\left(\frac{30\,\rho^{\chi}_{\rm ht}}{\pi^2 g_*^{\rm ht}}\right)^{1/4}\simeq 2.7 \times 10^{8} \,\text{GeV} \left(1+\frac{z_{\rm rad}}{\nu}\right)^{-3/4} \, \left( \frac{\Theta_{\text{ht}}}{10^{-8}}\right)^{3/4}  \left( \frac{H_{\rm kin}}{10^{11} \, \text{GeV}}\right)^{1/2}  \,,   
\end{equation}
with $g_*^{\rm ht}=106.75$ the Standard Model number of relativistic degrees of freedom at energies above ${\cal O}(100)$ GeV and $\rho^{\chi}_{\rm ht}=\rho^{\chi}(z_{\rm ht})=\rho^{\phi}(z_{\rm ht})$ the total energy-density of the scalar field at the end of the (re)heating phase. These results feed directly into constraints on inflationary model building. In particular, the minimal number of $e$-folds $N$ required to solve the flatness and horizon problems depends on Ricci reheating efficiency \eqref{eq:fittingTheta} and the virialization time \eqref{zradfit} \cite{Laverda:2023uqv},
\begin{equation} \label{eq:nefolds_inflation_2}
 N \simeq 63.3 + \frac14 \ln \left( \frac{r}{10^{-3}} \right) - \frac14 \ln \left( \frac{\Theta_{\text{ht}}}{10^{-8}} \right)\, + \frac14 \ln \left(1 + \frac{z_{\rm rad}}{\nu}\right)\,,
\end{equation}

\section{Gravitational wave production}\label{sec:grav}

The violent, non-equilibrium dynamics triggered by a HIPT act as a potent source of gravitational waves (GWs) \cite{Bettoni:2018pbl,Bettoni:2024ixe}. As symmetry is spontaneously broken after inflation, the resulting scalar field undergoes a tachyonic instability, followed by fragmentation and non-linear dynamics. This sequence of events—domain wall formation, their oscillations and collisions, and subsequent turbulent field evolution—gives rise to large time-varying quadrupole stresses, efficiently sourcing tensor perturbations in spacetime and generating a stochastic background of GWs.

Recent progress using classical lattice simulations has, for the first time, yielded a quantitative description of the GW spectrum produced during HIPTs \cite{Bettoni:2024ixe}. These simulations solve the full non-linear field dynamics coupled to the transverse-traceless components of the metric perturbations, allowing for direct extraction of the GW energy density as a function of frequency. The result is a broad yet sharply-peaked GW spectrum, with the peak determined by the typical size and evolution time of the inhomogeneities—roughly of sub-horizon scale at the time of the transition. A significant outcome of this work is the derivation of simple parametric fitting formulas for the main features of the spectrum, eliminating the need for repeated expensive simulations. Denoting the dimensionless peak momentum as $\kappa_p = k_p / H_{\rm kin}$ and the normalized peak amplitude as $\bar{\Omega}_{\rm GW,p} = \rho_{\rm GW} / |\rho_\chi|$ (both evaluated at the radiation time $z_{\rm rad}$), one finds \cite{Bettoni:2024ixe}
\begin{equation}
\kappa_p(\nu, \lambda) = \alpha_1 + \alpha_2 \nu\,, \quad \quad 
\ \bar{\Omega}_{\rm GW,p}(\nu, \lambda) = \left( \frac{H_{\rm kin}}{10^{10} \, \text{GeV}} \right)^2 \exp\left[ \beta_1 + \beta_2 \log \nu \right]\,,
\end{equation} 
with 
\begin{equation}
 \alpha_1 = -15.36 - 1.95n\,, \quad \quad \alpha_2 = 5.29 - 0.08n\,, \quad \beta_1 = -50.92 + 0.40 n \; , \hspace{5mm}
    \beta_2 = 7.80 + 0.55 n \,.
\end{equation}
 The total GW energy density fraction is similarly fitted by \cite{Bettoni:2024ixe}
\begin{align} 
\bar{\Omega}_{\rm GW}(\nu, \lambda) &= \left( \frac{H_{\rm kin}}{10^{10} \, \text{GeV}} \right)^2 \exp\left[ \gamma_1 + \gamma_2 \log \nu \right]\,, 
\end{align} with 
\begin{equation}
\gamma_1 = -50.50 + 0.40n\,,\quad\quad \gamma_2 = 7.70 + 0.55n\,. 
\end{equation}
The present-day GW signal can be obtained by redshifting these quantities through the cosmic expansion. The energy density today is given by \cite{Bettoni:2024ixe}
\begin{align} \Omega_{\rm GW,0} = 1.67 \times 10^{-5} h^{-2} \left( \frac{100}{g_*^{\rm ht}} \right)^{1/3} \bar{\Omega}_{\rm GW}\,, 
\end{align} 
with $h$ the dimensionless Hubble parameter $h\equiv H_0/(100 \,\textrm{km} \cdot s^{-1}\textrm{ Mpc}^{-1})$,  while the redshifted peak frequency reads
\begin{align}
f_{\rm p,0} \simeq 1.3 \times 10^9 \, \text{Hz} \cdot \frac{\kappa_p}{2\pi} \left( \frac{H_{\rm kin}}{10^{10}, \text{GeV}} \right)^{1/2} \left( \frac{\Theta_{\rm ht}}{10^{-8}} \right)^{-1/4} \sqrt{a_{\rm rad}}\,.
\end{align}
where $\Theta_{\rm ht}$ is the (re)heating efficiency (a function of $H_{\rm kin}$, $\lambda$, and $\nu$), defined as the ratio of the scalar to inflaton energy densities at the (re)heating time. The shape of the GW spectrum is well-captured by a broken power-law \cite{Bettoni:2024ixe}
\begin{align} \bar{\Omega}_{\rm GW}(f) = \bar{\Omega}_{\rm GW,p} \cdot \frac{(a + b)^c}{a \left(\frac{f}{f_p}\right)^{b/c} + b \left(\frac{f}{f_p}\right)^{-a/c}}^c\,,
\end{align} 
with fitted parameters $a = 3.00$, $b = 152.34 - 6.57\nu$, and $c = 105.85 - 4.79\nu$. This form mirrors similar expressions used in first-order phase transitions \cite{Lewicki:2020azd, Lewicki:2022pdb} and provides a flexible template for forecasting detectability. The frequencies involved typically span from MHz to GHz, determined by the scale of kination $H_{\rm kin}$. This high-frequency range lies beyond the reach of interferometers such as LIGO or LISA and is currently accessible only through resonant cavity experiments \cite{Gatti:2024mde, Aggarwal:2020olq}, though their sensitivities remain well below the constraints imposed by the effective number of relativistic of degrees of freedom during Big Bang Nucleosynthesis.
Importantly, the spectral shape from HIPTs is distinctive: a steep $f^3$ rise in the infrared (consistent with causality arguments \cite{Caprini:2009fx, Cai:2019cdl}), followed by a rapid UV fall-off whose slope depends on the duration and strength of the non-linear phase. This differs markedly from the nearly scale-invariant background from inflation or the broader spectra from bubble collisions, providing a potential observational discriminant. Moreover, the amplitude of the signal carries information about the fraction of energy in the field and the timing of the transition relative to inflation’s end. 

\section{The Standard Model Higgs as a case study}\label{sec:Higgs}

A particularly well-motivated example of HIPT involves the Standard Model (SM) Higgs field. The Higgs is present in the early Universe and, being a scalar field, it is susceptible to the kind of curvature-induced instability discussed above.~\footnote{It is important to stress that, despite the involvement of a non-minimal coupling between the Higgs field and curvature in both cases, the scenario studied here is fundamentally different from Higgs inflation \cite{Rubio:2018ogq}. In Higgs inflation, the field itself plays the role of the inflaton and must begin with a large, Planck-scale amplitude to drive the accelerated expansion. By contrast, the present framework treats the Higgs as a subdominant spectator field whose initial value is close to zero within the inflating patch. This distinction is not merely technical — it profoundly alters both the field dynamics and the relevant consistency and stability considerations. In particular, for successful configurations, the present scenario is not dramatically sensitive to potential threshold effects emerging at the cut-off scale of the SM non-minimally coupled to gravity \cite{Bezrukov:2014ipa,Rubio:2015zia}.} In fact, the SM Higgs potential is known to have a metastability issue at high energies: due to renormalization group running, the Higgs self-coupling may become negative at scales around $10^{10}$–$10^{15}$ GeV, creating a shallow barrier and a second, lower vacuum at enormous field values. This means our electroweak vacuum could be a false vacuum that might decay given sufficient energy or fluctuations. During inflation, if the Hubble scale is high, quantum fluctuations of the Higgs could in principle kick it over the barrier into the catastrophic true vacuum. However, adding a non-minimal coupling for the Higgs can stabilize it during inflation by giving it a large positive mass term from the curvature. Many inflationary scenarios assume such a coupling to prevent vacuum decay, see e.g. \cite{ Herranen:2014cua, Herranen:2015ima,Laverda:2024qjt}

\begin{figure}
    \centering
  \includegraphics[width=\textwidth]{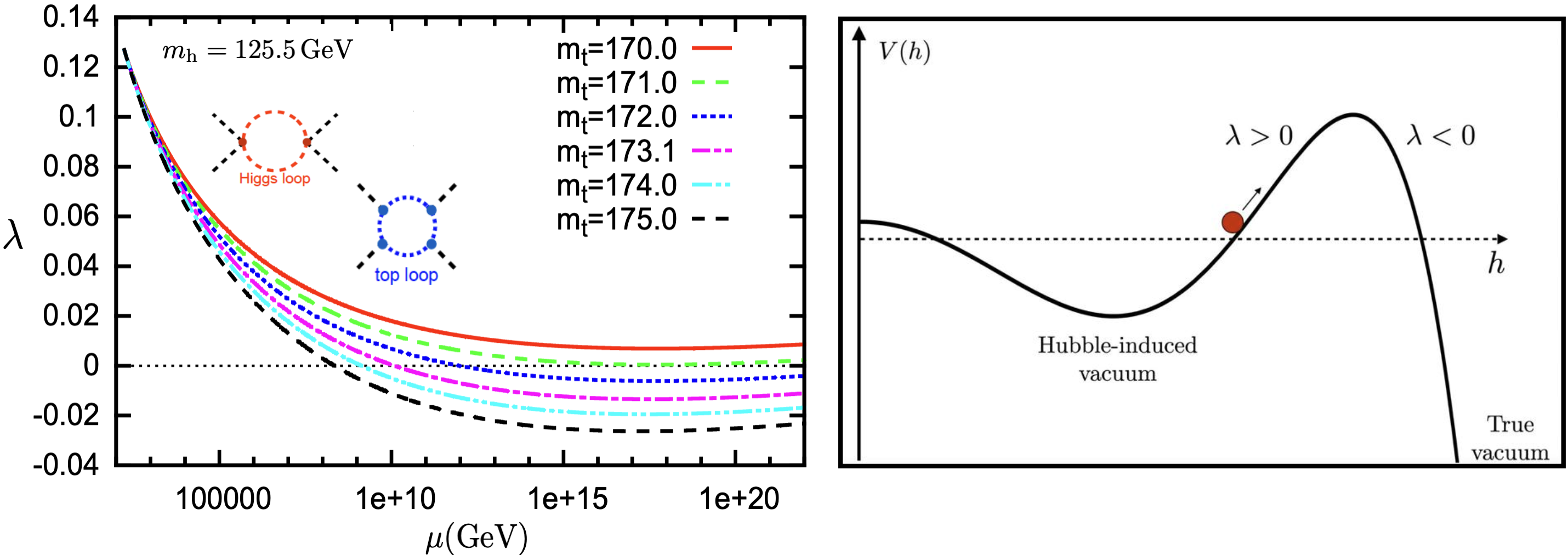}
    \caption{Left: Running of the Higgs quartic coupling $\lambda$ as a function of the renormalization scale $\mu$ for different values of the top quark mass. Taken from \cite{Bezrukov:2014ipa}. Right: Schematic Higgs potential showing the emergence of a metastable Hubble-induced vacuum and a deeper true vacuum at large field values, illustrating the consequences of the running coupling.}
    \label{fig:higgs_running_schematic}
\end{figure}

The irony is that after inflation, during a kination phase, this same coupling term becomes a negative mass term when $R$ flips sign and can drive the Higgs towards the instability region, cf.~Fig.~\ref{fig:higgs_running_schematic}. In other words, the Higgs field can undergo a HIPT of its own \cite{Laverda:2024qjt}. The crucial question is: does it overshoot into the true vacuum, which would be disastrous, or does it settle into a large vev corresponding to the metastable Hubble-induced part of the potential? Detailed analyses show that avoiding a collapse of the electroweak vacuum imposes a stringent upper bound on the inflationary Hubble scale or equivalently on the energy scale of inflation, given a particular $\xi$ and assuming the known SM parameters. Essentially, $H_{\rm kin}$ must be low enough, or $\xi$ small enough, that the energy injected into Higgs fluctuations remains below the height of the potential barrier separating the false vacuum from the true one. If this condition is satisfied, the Higgs will not roll into the catastrophic vacuum during the kination era – the symmetry breaking induced by $R<0$ will only take it partway to the vicinity of the barrier but not over it. Note that this is indeed a delicate balance as the bound also depends on the precise height of the Higgs potential barrier, which in turn depends sensitively on the top quark Yukawa coupling influencing the running of $\lambda$. In particular, for a Higgs mass fixed to its current best-fit central value, $m_h=125.20\pm0.11 \textrm{ GeV}$, and a strong coupling constant value $\alpha_s=0.1180$ \cite{ParticleDataGroup:2024cfk}, the complete numerical three-loop $\overline{\rm MS}$ running of the Higgs self-coupling \cite{BezrukovNotebook} within the energy scale range $\mu\in[10^4 \textrm{ GeV}, \, 10^{23} \textrm{ GeV}]$ admits a parametric form\footnote{This parametrisation reproduces the full numerical running with an accuracy better than 1\% across the considered range of renormalisation scales and for top quark masses within $m_t\in[169 \textrm{ GeV} \; ,175 \textrm{ GeV}]$ \cite{Laverda:2025pmg}.} \cite{Laverda:2025pmg}
\begin{equation}
    \lambda_{\rm fit}(\mu)=\lambda_0+c_1\log \left( \frac{\mu}{\rm GeV} \right) +c_2\log^2 \left( \frac{\mu}{\rm GeV} \right) +c_3\log \left( \log\left(\frac{\mu}{\rm GeV}\right) \right)\,,
    \label{eq:running_fit}
\end{equation}
with the coefficients $\lambda_0$, $c_1$, $c_2$, and $c_3$ depending on the top quark mass as
\begin{align}
    \lambda_0(m_t)&=7.31965 -9.13428\times 10^{-3}\times \frac{m_t}{\rm GeV}+2.94355\times 10^{-4}\times \left(\frac{m_t}{\rm GeV}\right)^2 \; ,\nonumber \\
    c_1(m_t)&=1.98329\times 10^{-1} -2.34181\times 10^{-3}\times \frac{m_t}{\rm GeV}+7.05298\times 10^{-6}\times \left(\frac{m_t}{\rm GeV}\right)^2 \; ,\nonumber \\
    c_2(m_t)&=-1.4835\times 10^{-4} + 7.88043\times 10^{-7}\times \frac{m_t}{\rm GeV} \; ,\nonumber \\
    c_3(m_t)&=-4.19967 + 5.30679\times 10^{-2}\times \frac{m_t}{\rm GeV}-1.70896\times 10^{-4}\times \left(\frac{m_t}{\rm GeV}\right)^2 \; .
\end{align}
Interestingly, current measurements of the top quark’s pole mass around ~171 GeV \cite{CMS:2019esx} are such that the SM is just barely metastable, and slightly lower top mass values make the vacuum absolutely stable. In the context of the Hubble-induced transition, a slightly lower top quark mass (within the experimental uncertainty) is favourable, as it raises the barrier and makes it easier to fulfill the stability condition. Thus, the requirement of electroweak vacuum stability during a Hubble-induced transition translates into a constraint on fundamental parameters: it links the inflationary scale, the Higgs-curvature coupling, and the top quark mass. Any viable cosmic history that includes a kination era must respect these bounds to ensure our vacuum’s survival \cite{Laverda:2024qjt}. 

If the SM Higgs field undergoes a high-scale instability phase transition (HIPT) safely---falling into the electroweak vacuum's metastable valley without transitioning to the deeper true vacuum---it can itself act as the reheaton \cite{Laverda:2024qjt}. In this scenario, referred to as \textit{Higgs reheating} in what follows, the energy stored in the Higgs oscillations is directly transferred to the Standard Model degrees of freedom, leading to a rather fast thermalization. For this mechanism to be effective, the inflationary Hubble scale must exceed a critical value, $H_{\rm kin} \gtrsim 10^{5.5}~\mathrm{GeV}$, as a lower scale would render Higgs-induced particle production insufficient to trigger a radiation-dominated era before Big Bang Nucleosynthesis. On the other hand, vacuum stability imposes an upper limit on $H_{\rm kin}$, implying that successful Higgs reheating can only occur within a specific intermediate range. Remarkably, this window is viable for slightly lower top quark masses. In this minimalistic framework, the Higgs field fulfills two roles: maintaining vacuum stability during inflation through its non-minimal coupling to gravity, and reheating the Universe afterward without invoking additional scalar fields or external reheating mechanisms~\cite{Laverda:2024qjt}.



\begin{figure}[tb]
\centering
\includegraphics[width=0.85\textwidth]{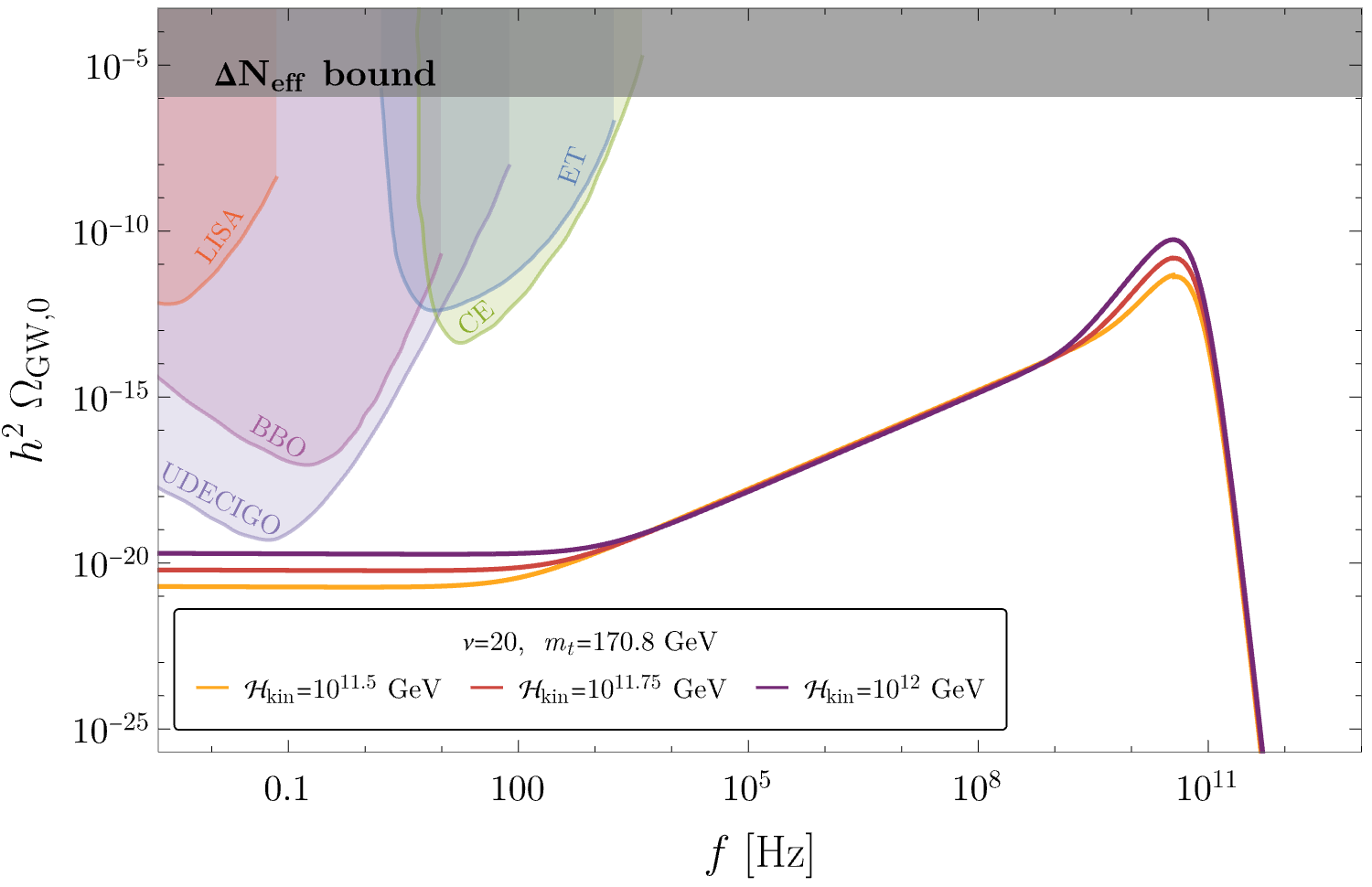}
\hspace{5mm}
\caption{Spectral shape of the SGWB produced by a HIPT for various symmetry-breaking scales, shown alongside the projected sensitivities of upcoming gravitational wave detectors: Laser Interferometer Space Antenna (LISA) \cite{amaroseoane2017laserinterferometerspaceantenna, Robson:2018ifk}, Big Bang Observer (BBO) \cite{Crowder:2005nr, Corbin:2005ny}, UltimateDECIGO \cite{Seto:2001qf, Yagi:2011wg, Kawamura:2020pcg}, Einstein Telescope (ET) \cite{Punturo:2010zz, Branchesi:2023mws}, and Cosmic Explorer (CE) \cite{LIGOScientific:2016wof, Reitze:2019iox}. The non-minimal coupling is fixed at $\nu=20$, and the top quark mass is set to $m_t = 170.8 \textrm{ GeV}$ to guarantee absolute stability of the Higgs vacuum. Taken from \cite{Laverda:2025pmg}. }
\label{fig:GW_absolute stability}
\end{figure}

\begin{figure}[tb]
    \centering
        \includegraphics[width=0.47\textwidth]{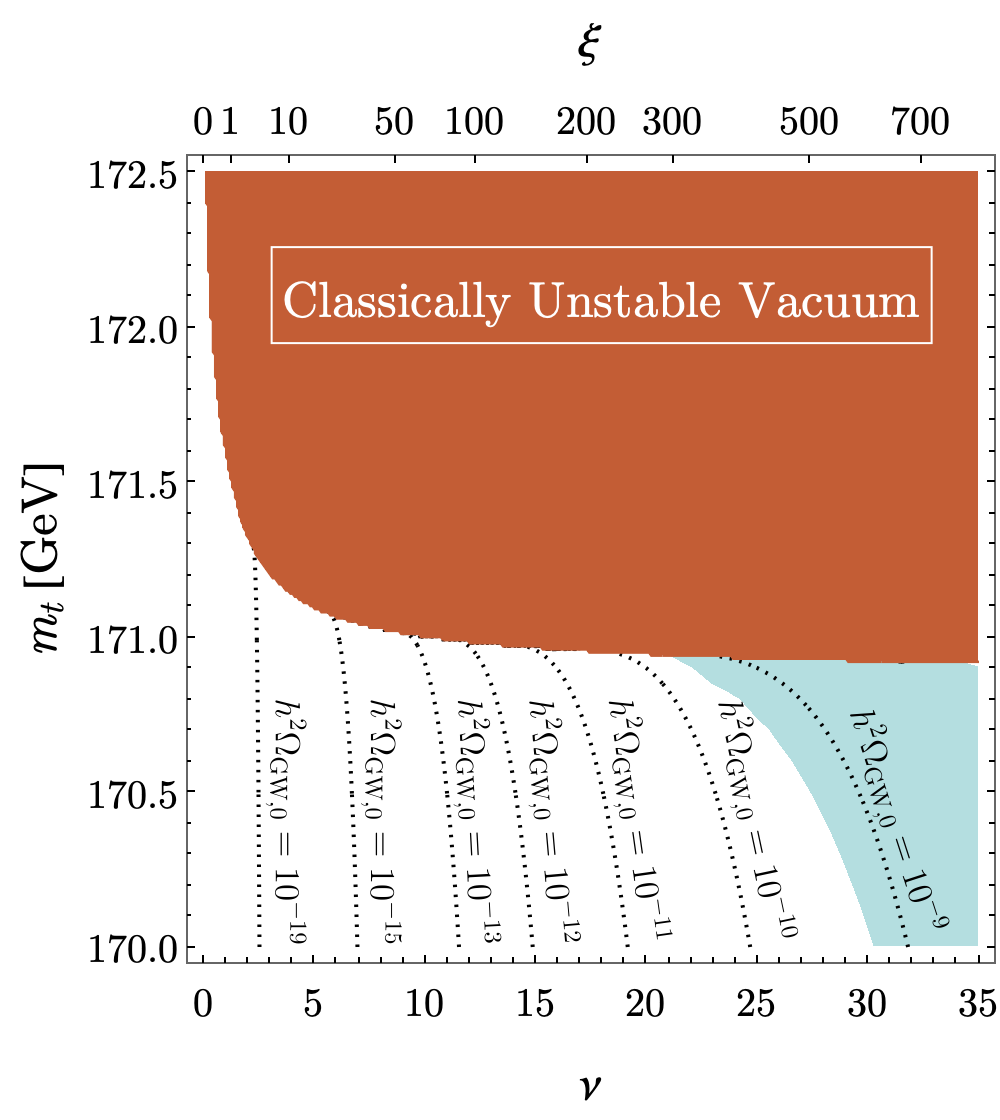}
        \hspace{1mm}
        \includegraphics[width=0.47\textwidth]{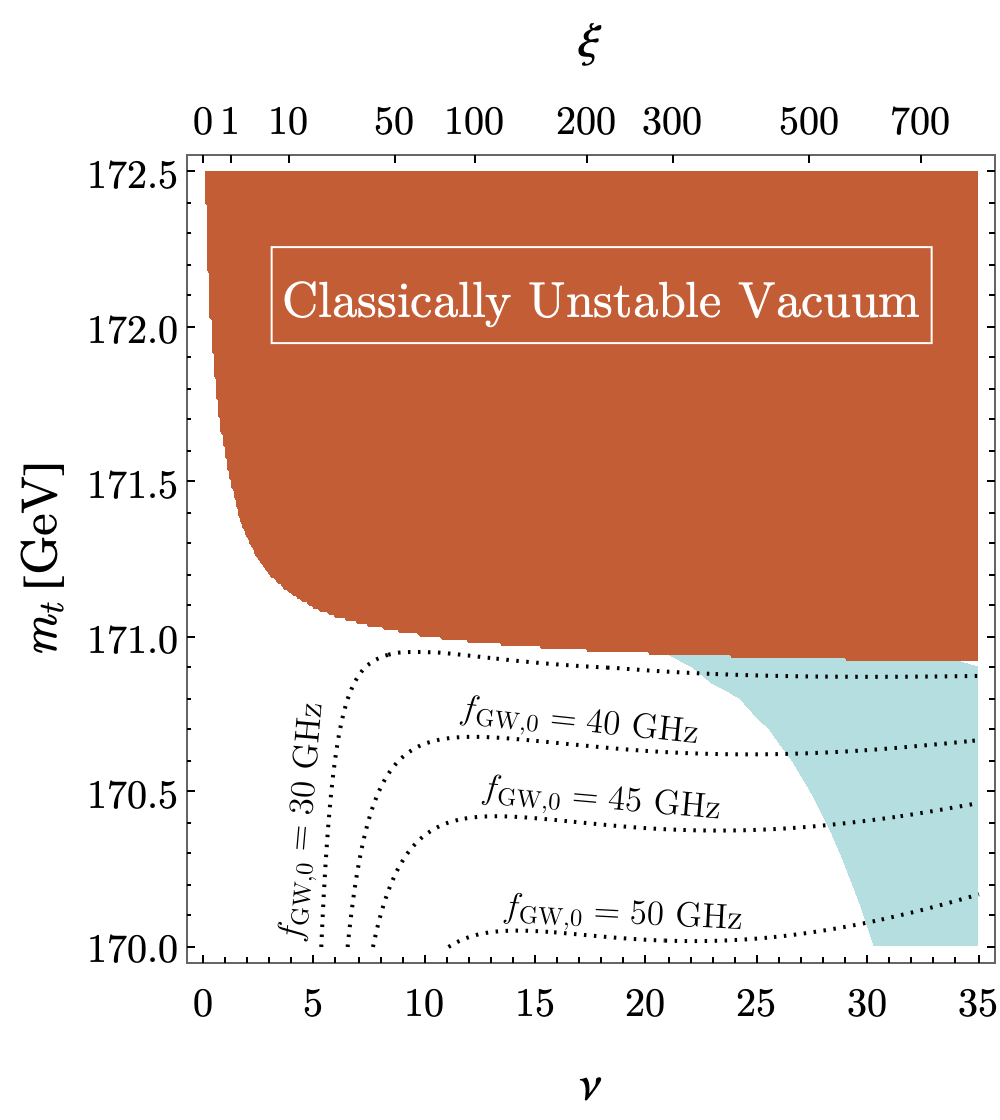}
        \hspace{4mm}
        \caption{Present-day gravitational wave energy density and peak frequency as functions of the top-quark mass and the non-minimal coupling parameter, for a fixed phase transition scale \( \mathcal{H}_{\rm kin} = 10^{12} \, \text{GeV} \). The red regions correspond to classically unstable configurations, while the light-blue areas are excluded for violating the 1\% upper bound relative to the gravitational theory's cutoff scale \cite{Bezrukov:2010jz}. Taken from \cite{Laverda:2025pmg}.
} \label{fig:res_cav_top_mass}
\end{figure}

An especially intriguing aspect of the Higgs spectator field case is the predictive power it offers for observational signals, notably gravitational waves. The dynamics here are essentially the same as described in general, so one expects a stochastic gravitational-wave background from the Higgs’s tachyonic oscillations and domain network. The difference is that now the properties of that GW background are directly related to known Standard Model parameters. For example, the peak frequency and amplitude of the GW signal can be tied to the inflationary Hubble scale, the Higgs’ self-coupling, and the Higgs-curvature coupling $\xi$. Even the top quark Yukawa coupling plays a role in determining the exact shape of the Higgs potential at high scales, and thus influences the details of the phase transition and its gravitational-wave output, cf.~Fig.~\ref{fig:res_cav_top_mass}.  
One can consider three different possible high-scale scenarios: (i) the Higgs is absolutely stable up to Planckian energy (which could happen if the top mass is low); (ii) the Higgs is metastable as in the SM (with a barrier and deeper true vacuum, but the transition does not actually take place thanks to $\xi$ stabilization); or (iii) new physics intervenes just beyond the SM instability scale, effectively modifying the Higgs potential. Each case can lead to slightly different outcomes in the Hubble-induced transition. For instance, in case (i) there is actually no true vacuum to worry about – the Higgs would just undergo a symmetry-breaking to some large vev and oscillate there, which might produce a somewhat different GW spectrum than case (ii) where the field approaches a dangerous region and then retreats. Nevertheless, all these scenarios produce a significant gravitational-wave signal from the Higgs field’s dynamics. The observational prospects were recently highlighted \cite{Laverda:2024qjt}: if a peaked stochastic GW background were detected in the future, and if it were accompanied by the distinctive high-frequency tail expected from an inflationary origin (amplified primordial tensor modes), one could effectively link the signal to specific features of the Higgs potential. In other words, gravitational waves could become a probe of the Higgs field behaviour at energies around $10^{11}$ GeV – far beyond the reach of any collider, but accessible through the imprint left on spacetime ripples. Such a detection would be revolutionary: it would simultaneously verify a non-standard (re)heating mechanism (Ricci/Higgs reheating) and reveal insights about the Higgs potential (and thus the top quark Yukawa, etc.) at otherwise inaccessible scales. It’s important to note that this gravitational-wave production does not rely on the Higgs vacuum decaying via quantum tunnelling. Unlike some proposals where one might get a GW burst from an actual first-order vacuum decay (which in the SM Higgs case would be fatal to our Universe!), here the GWs are generated by classical field dynamics in a perfectly consistent history where the electroweak vacuum remains intact. This makes the scenario safer and more predictive, since it depends on well-defined classical physics rather than exponentially rare quantum events.

\section{Conclusions and outlook}\label{sec:conc}

HIPTs represent a compelling bridge between high-energy particle physics and early-Universe cosmology. In this framework, the simple fact of a changing Hubble rate can act as the trigger for symmetry breaking, offering an elegant alternative to thermally driven phase transitions. We have outlined the general mechanism by which a non-minimally coupled scalar field can undergo spontaneous symmetry breaking during a post-inflationary kinetic era, surveying the rich sequence of consequences: from the formation and eventual destruction of domain walls and other defects, to the (re)heating of the Universe via curvature-driven particle production (Ricci reheating), and the generation of a stochastic background of gravitational waves. These phenomena lead to potentially observable signatures, most notably a stochastic gravitational wave spectrum, which future detectors might be able to probe. Moreover, when applying this general idea to the Standard Model Higgs, we find a scenario that ties cosmological history to measurable Standard Model parameters like the top quark mass. The requirement of vacuum stability under Hubble-induced symmetry breaking and the demand for successful Higgs reheating together single out a predictive parameter window that could be tested by both collider experiments (refining the top/Higgs parameters) and cosmic surveys (searching for primordial gravitational waves or other relics).

Looking ahead, there are several avenues for further investigation. On the theoretical side, one may explore Hubble-induced transitions in other frameworks “beyond the Standard Model” – for instance, pseudoscalar axion fields with non-minimal couplings, or symmetry breaking in grand unified theories during kination, etc. Each may have its own defect structures (strings, monopoles) and observable consequences. The impact on baryogenesis and dark matter production also merits study: since Hubble-induced transitions provide out-of-equilibrium conditions and abundant particle production, they could offer new ways to generate the baryon asymmetry \cite{Bettoni:2018utf} or dark matter relic density. Interestingly, they could offer an explanation of DESI results \cite{DESI:2024mwx}, motivating a late-time generation of neutrino masses \cite{Goertz:2024gzw}, or play also a role in non-thermal \textit{first-order phase} transitions \cite{Kierkla:2023uzo,Mantziris:2024uzz} 

On the observational side, improving sensitivity to a stochastic gravitational-wave background across a broad range of frequencies will be crucial. If a signal consistent with a HIPT were detected, it would not only confirm the existence of a non-standard post-inflation epoch, but also open a window into the particle physics of that epoch. Conversely, even the absence of a signal in certain frequency ranges can constrain the parameter space of these models (for example, ruling out very high (re)heating scales that would have produced an observable GW background). In summary, HIPTs enrich the paradigm of the early Universe by linking cosmological dynamics with particle physics in novel ways. They provide a plausible narrative for how our Universe could smoothly connect inflation to the hot Big Bang, all while potentially leaving behind distinctive imprints for us to detect. 

\section{Acknowledgments}

J.~R. is supported by a Ram\'on y Cajal contract of the Spanish Ministry of Science and Innovation with Ref.~RYC2020-028870-I. This research was further supported by the project PID2022-139841NB-I00 of MICIU/AEI/10.13039/501100011033 and FEDER, UE. G.~L. acknowledges support from a fellowship provided by ``la Caixa” Foundation (ID 100010434) with fellowship code LCF/BQ/DI21/11860024, as well as the support of FCT through the grant with Ref.~2024.05847.BD. G.~L. acknowledges also the financial support from FCT to the Center for Astrophysics and Gravitation-CENTRA, Instituto Superior Técnico, Universidade de Lisboa, through Project No.~UIDB/00099/2020. 

\bibliographystyle{JHEP}
\bibliography{bibliography}
\end{document}